\documentclass[12pt]{article} 

\usepackage{PRIMEarxiv}
\usepackage[utf8]{inputenc} 
\usepackage[T1]{fontenc}    

\usepackage{hyperref}       
\usepackage{url}            
\usepackage{booktabs}       
\usepackage{amsfonts}       
\usepackage{nicefrac}       
\usepackage{microtype}      
\usepackage{graphicx}       
\usepackage{lipsum}         
\usepackage{times}          
\usepackage{setspace}       
\usepackage{amsmath}
\usepackage{relsize}   
\usepackage{anyfontsize}       
\usepackage[font=small,labelfont=bf]{caption} 

\renewcommand\normalsize{\fontsize{11}{15}\selectfont}

\title{\textbf{Ultrafast pulse propagation time-domain dynamics in dispersive one-dimensional photonic waveguides}}

\author{
  Ahmet Oguz Sakin$^{1}$, Ali Murat Demirtas$^{1}$, Hamza Kurt$^2$, Mehmet Unlu$^{1*}$ \\
  $^1$Department of Electrical and Electronics Engineering, TOBB University of Economics and Technology, Ankara, Turkey \\
  $^2$School of Electrical Engineering, Korea Advanced Institute of Science and Technology (KAIST), Daejeon, Republic of Korea \\
  \textbf{*Corresponding author:} \texttt{munlu@etu.edu.tr} \\
}
\makeatletter
\renewcommand{\maketitle}{
  \begin{center}
    {\LARGE \textbf{\@title} \par}
    \vspace{1em}
    {\large \@author}
  \end{center}
}
\makeatother
\begin{document}
\maketitle
\begin{abstract}
Ultrafast pulses, particularly those with durations under 100 femtoseconds, are crucial in achieving unprecedented precision and control in light-matter interactions. However, conventional on-chip photonic platforms are not inherently designed for ultrafast time-domain operations, posing a significant challenge in achieving essential parameters such as high peak power and high temporal resolution. This challenge is particularly pronounced when propagating through dispersive integrated waveguides, unlike classical applications where dispersion is typically near-zero or linear. In addressing this challenge, we present a design methodology for ultrafast pulse propagation in dispersive integrated waveguides, specifically focused on enhancing the time-domain characteristics of one-dimensional grating waveguides (1DGWs). The proposed methodology aims to determine the optimal structural parameters for achieving maximum peak power, enhanced temporal resolution, and extended pulse storage duration during ultrafast pulse propagation. To validate this approach, we design and fabricate two specialized 1DGWs on a silicon-on-insulator (SOI) platform. A digital finite impulse response (FIR) model, trained with both transmission and phase measurement data, is employed to obtain ultrafast time-domain characteristics, enabling easy extraction of these results. Our approach achieves a 2.78-fold increase in peak power and reduces pulse broadening by 24\%, resulting in a smaller sacrifice in temporal resolution. These results can possibly pave the way for advanced light-matter interactions within dispersive integrated waveguides.
\end{abstract}


\section{Introduction} 
In the perpetually advancing field of ultrafast photonics, the advent of Photonic Integrated Circuits (PICs) employing femtosecond laser pulses represents a significant technological advancement. This progress has wide-ranging implications across various fields, such as quantum photonics \cite{yanagimoto2022temporal, williams2024ultrashort}, photonic neural networks \cite{wang2022integrated}, terahertz waveform generation \cite{herter2023terahertz, kang2024frequency, takano2024frequency}, and sensing \cite{guo2018mid}. As these applications become more widespread, fine-tuning the structural parameters of PICs becomes crucial to meet the specific requirements of ultrafast photonic systems. This is particularly important when utilizing femtosecond-duration pulses, which are further complicated by the considerable increase in pulse intensity distortion and temporal broadening during propagation \cite{hickstein2019self}. The propagation of optical Gaussian pulses in linear dispersive media is modeled by the one-dimensional Schrödinger equation, as shown in Equation 1 \cite{agrawal2013nonlinear}.

\begin{equation}
\label{deqn_ex1a22}
E(z,t) =  \, A_0 \frac{T_0}{\sqrt{T_0^2 + i \frac{T_0^2 z}{L_D}}}  \times \exp \left( -\frac{(t - \frac{z}{v_g})^2}{2\left(T_0^2 + i \frac{T_0^2 z}{L_D}\right)}\right) 
\end{equation}

\noindent Here, \( E(z,t) \) represents the output electric field envelope of the pulse, where \( A_0 \) denotes the initial amplitude, and \( T_0 \) corresponds to the initial pulse width. The dispersion length, represented by \( L_D = \frac{T_0^2}{|\beta_2|} \), is determined by the group velocity dispersion (GVD) parameter \( \beta_2 \), and the pulse's group velocity is denoted by \( v_g \). This equation demonstrates that pulse broadening and the decay of peak intensity are significantly correlated with the dispersion length. When comparing Gaussian pulses with durations of 90 fs and 10 ps, the dispersion length indicates that the broadening and intensity decay experienced by a 10 ps pulse over a distance of 1 meter are equivalent to those experienced by a 90 fs pulse over just 81 µm. These limitations for near-zero or linear dispersion media become even more pronounced when ultrafast signals propagate through highly dispersive media. It is also important to note that, although not included in Equation 1, nonlinear effects and the third-order dispersion (TOD) coefficient \( \beta_3 \) notably influence the asymmetry and chirping of femtosecond-duration signals \cite{agrawal2013nonlinear,gaafar2024femtosecond}.

In recent times, there has been a discernible surge of interest in the engineering of slow light phenomena using dispersive integrated waveguides aimed at extending light-matter interaction, improving optical nonlinearity, and establishing a genuine true-time-delay mechanism \cite{krauss2008we,boyd2011material}. One-dimensional grating waveguides (1DGWs), an archetype of highly dispersive structures, have attracted considerable interest due to their streamlined design and reduced dimensionality. This design simplifies the fabrication process by reducing the etched surface area that interacts with the optical mode \cite{chung2018chip}. Furthermore, when configured with an appropriate Bloch mode profile, 1DGWs are well-suited for high-density integration, effectively minimizing crosstalk in densely populated chip arrangements \cite{wang2015low,chung2022ultra}. Nonetheless, research endeavors focused on achieving slow light using 1DGWs typically target the generation of a sharp dispersion profile in proximity to the band edge \cite{chung2018chip,wang2018continuously}. This approach often restricts bandwidth accommodation due to the time-bandwidth product, resulting in significant pulse distortion and posing a notable limitation in ultrafast photonic applications \cite{tsakmakidis2021topological, bao2011flat}. Despite these problems, this design method continues to be widely applied to almost all dispersive photonic components. Consequently, a significant challenge persists in achieving optimal ultrafast pulse propagation in dispersive integrated waveguides, particularly for 1DGWs to fully exploit their aforementioned advantages. 

\begin{figure*}[!t]
\centering
\includegraphics[height=2.5in,width=6.5in]{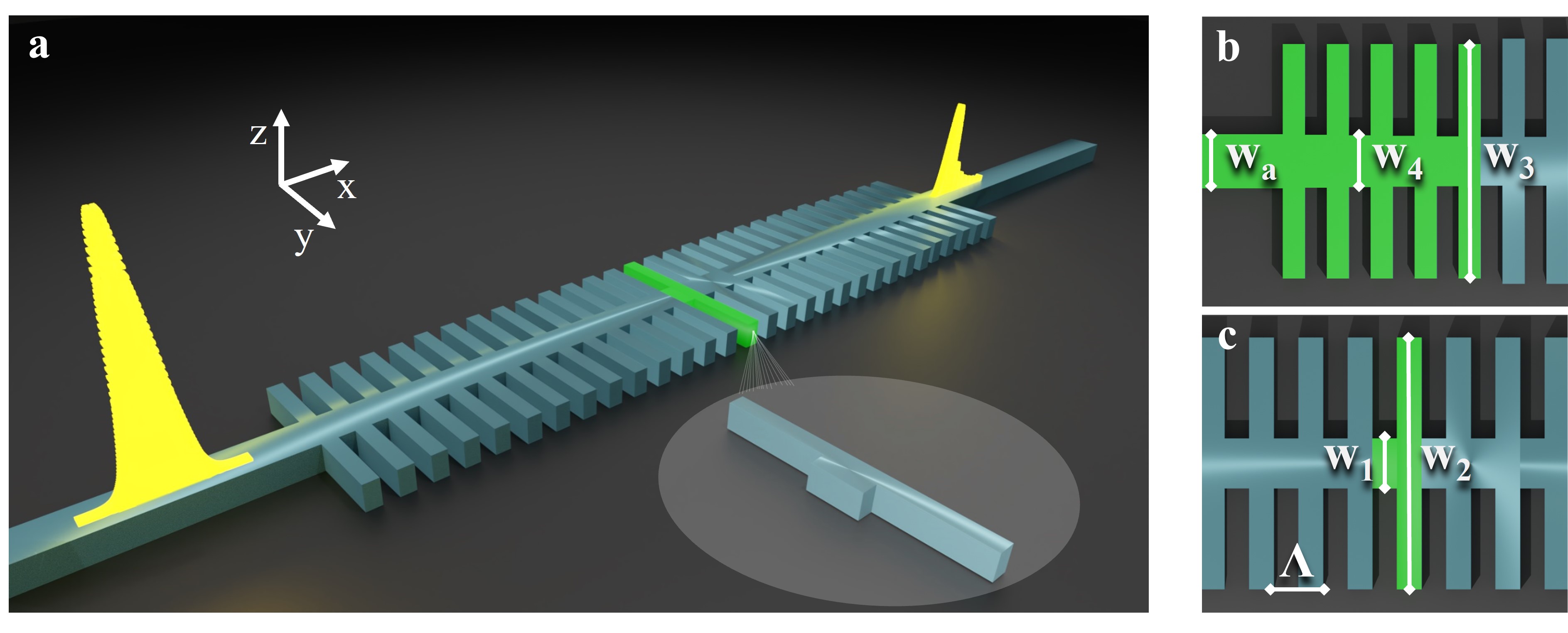}
\label{s1}
\caption{(a) Schematic representation of the 1DGW structure, (b) top view of the transition region, (c) top view of the periodic topology, highlighting a single period.}
\end{figure*}

In light of these challenges, one prospective approach for mitigating the limitation of 1DGWs entails employing a narrower corrugation width configuration \cite{yu2022integrated, choi2022picosecond}. This configuration yields a Bloch mode profile with greater symmetry and a substantially lower group index. Nevertheless, it necessitates a greatly extended structural length to achieve the desired light-matter interaction, thereby imposing constraints on densely packed chip layouts. In an alternative approach, an increased bandwidth has been achieved while maintaining fixed group index values by utilizing optimized minuscule unit squares that introduce interruptions in the corrugated grating structure \cite{jiang2022exploring,hao2019increasing}. However, a critical aspect often overlooked in optimizing these minuscule unit squares in 1DGWs is the analysis of temporal domain characteristics. Minor variations in bandwidth increment in these studies, despite their impact on the decay in pulse intensity distortion and pulse broadening, are insufficient to prevent ultrafast time-domain problems. This issue stems from an excessive focus on frequency domain optimization. Moreover, imperfections such as nonlinear effects, higher-order dispersion, and variations in temporal coupling coefficients within the transition region are often overlooked. Therefore, the design of structures for time-domain studies, particularly in ultrafast pulse propagation through highly dispersive media, should prioritize parameters that are critical for time-domain applications.

In this paper, we present a design methodology for enhancing time-domain dynamics during ultrafast pulse propagation in dispersive integrated waveguides, focusing on 1DGWs. The method optimizes 1DGW structural parameters using a figure of merit (FoM) that maximizes the product of normalized peak intensity and time delay through ultrafast pulse propagation in highly dispersive media. To validate its effectiveness, we design and fabricate two specialized 1DGW structures: one optimized for high group index and the other for ultrafast time-domain characteristics. For a simplified and accurate extraction of the ultrafast time-domain response, we employ a digital FIR approach trained on fully measured data. The following sections outline the proposed methodology and present the results.



\section{Structure and Design}
\subsection{Determining Range of Physical Dimensions}

Figure 1-(a) to (c) presents an overview of the 1DGW design. The structural framework is founded upon an SOI substrate, employing a low-resistivity wafer composed of silicon with a depth of 220 nm in conjunction with a buried oxide (BOX) layer measuring 2 $\mu$m  in thickness. The structure of the 1DGW entails a systematic modulation of the effective refractive index along the optical mode's propagation axis. The modulation of effective indices within the gratings can be achieved through various sidewall modulation techniques, such as square, sawtooth, triangular, or sinusoidal patterns, etc. \cite{cheng2019apodization}. However, these apodization methods introduce undesirable localized variations in the effective refractive index of the inner waveguide and corrugated part of the grating structure. This can lead to phase noise and an increase in output pulse distortion. Hence, we elect to utilize a consistent index modulation denoted by $\Delta n$ as the configuration of non-apodized gratings.

\begin{figure}[!ht]
\centering
\includegraphics[width=3in]{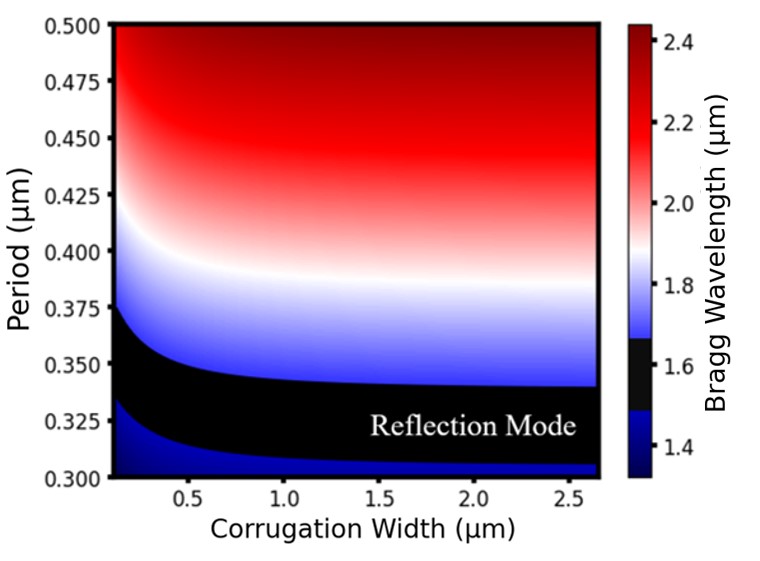}
\caption{Calculation of theoretical Bragg wavelength based on the values of the grating period and the width of the corrugation. The segment highlighted in the black region indicates the range of parameters conducive to the reflection mode within the 1.5-1.65 µm wavelength range.}
\label{s2}
\end{figure}

\begin{figure*}[!ht]
\centering
\includegraphics[width=6.5in]{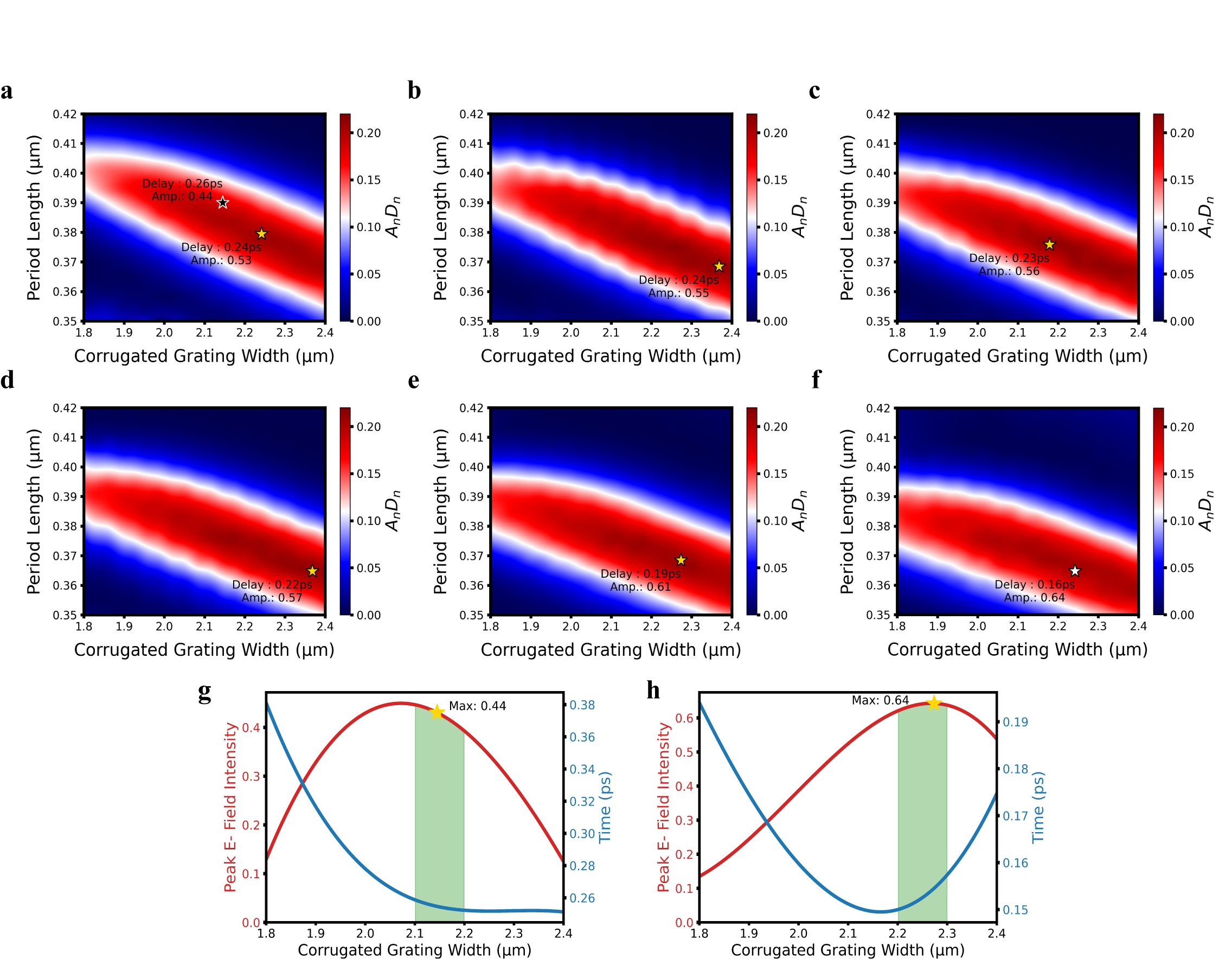}
\caption{The investigation of 1DGW parameters aims to achieve optimal time delay and peak intensity. In graphs (a) to (f), the x-axis represents the corrugated grating width, and the y-axis corresponds to the period width. These graphs reflect different settings of the $w_1$ parameter: 350 nm, 370 nm, 390 nm, 400 nm, 420 nm, and 450 nm, respectively. White and black stars indicate the points that have been chosen to demonstrate the effectiveness of the method, while gold stars mark the numerically optimal points identified for each parameter set. Graphs (g) and (h) show the peak intensity and time delay for the selected topologies (denoted by the black star from (a) and the white star from (f)), plotted against the corrugated grating width. Both graphs use a fixed period and $w_1$ parameter, illustrating the relationship between the corrugated width and the corresponding characteristics.}
\label{s3}
\end{figure*}

Due to the Fabry-Pérot cavity mechanism, Equation (2) describes the phase-matching condition between the transmitted and reflected modes. This equation is highly suitable for theoretically defining the limits of the structural parameters.

\begin{equation}
\label{deqn_ex1a}
\lambda_B = (n_{\text{eff1}} + n_{\text{eff2}}) \Lambda
\end{equation}

\noindent Here, $\lambda_{B}$ represents the Bragg wavelength, $\Lambda$ denotes the grating period, and $n_{eff1}$ and $n_{eff2}$ are the effective refractive indices of the inner waveguide ($w_1$) and the corrugated part ($w_2$) of the structure, respectively. Increasing the corrugation width ($\Delta w$), representing the difference between the corrugated and inner waveguide widths, enables intentional engineering to achieve a wider operational wavelength range \cite{liu2023chip}. Therefore, selecting the maximum limits for the parameter range of corrugation width ($\Delta w$) ensures the inclusion of structures with high bandwidth values in the process, potentially covering points where temporal signal distortion is minimized. When selecting the maximum limits of corrugation width, it is essential to consider a range that addresses the high coupling coefficient between inner waveguides and corrugated gratings. Additionally, increasing the corrugation width value can pose difficulties in the design of the transition from strip waveguide to 1DGW, as explained later in this section. In light of the difficulties identified, the inner waveguide of the grating, denoted as $w_1$, consistently adopted a value at the optimally acceptable range, specifically between 0.35 and 0.45 $\mu m$. When determining the parameter ranges for the corrugated grating width ($w_2$), and the periodicity ($\Lambda$) using Bragg wavelength calculation, the value of $w_1$ is fixed at 350 nm to satisfy the maximum corrugation width criterion. The Bragg wavelength is theoretically modeled using Equation (2), which is based on the structural parameters outlined in Figure 2. Within the spectrum defined by the Bragg wavelength, light experiences significant reflection. Therefore, to achieve constructive interference within the 1.5 - 1.65 $\mu m$ wavelength interval, the corrugated grating width ($w_2$) and periodicity ($\Lambda$) should exceed those corresponding to the reflected regions, as highlighted in the black areas of Figure 2. This approach helps prevent the emergence of reflection modes. Similarly, choosing higher periodicity values can lead to the operational range of transmission modes shifting to wavelengths exceeding the intended spectral range. Consequently, the corrugated grating and periodicity values should be selected within the ranges of 1.8-2.4 $\mu$m and 0.35-0.42 $\mu$m, respectively.

The final element to consider in the 1DGW configuration is the design of a high-bandwidth transition zone. Owing to the group-velocity mismatch between the strip waveguide and the 1DGW, there is a noticeable reduction in the butt-coupling efficiency between these structures. Rather than using a linear or adiabatic taper, which is insufficient for supporting slow-light coupling to a higher-band grating waveguide, an anti-Fresnel reflection method-based step taper approach is adopted. This approach enhances energy transfer efficiency, reduces loss, and minimizes oscillations \cite{zhao2017efficient}. It entails tapering the width in five distinct periods, denoted by $w_4$, which gradually narrows from the width of the strip waveguide ($w_a$) to the inner waveguide width of the 1DGW ($w_1$). Incorporated into the transition zone design, the parameter $w_3$— representing the toothed segment— is set to be uniformly 100 nm lower than the $w_2$ value. 

\subsection{Determining Physical Dimensions}

The system is excited using a Gaussian signal with a full width at half maximum (FWHM) of 90 femtoseconds (fs) at a wavelength of 1550 nm. The unit length, comprising mainly 50 periods of the structure, is kept constant to obtain precise pulse storage durations. The analysis is conducted using the three-dimensional finite-difference time-domain (3D-FDTD) solver within the defined parameter ranges. The ultrafast time-domain pulses, obtained at the end of the unit length, are then characterized. To evaluate the characterized time-domain results in relation to peak power, temporal resolution, and pulse storage duration, we employ metrics that directly correspond to these parameters. Specifically, we assess peak electric field (E-field) intensity to represent peak power, pulse broadening to capture the sacrifice of temporal resolution, and time delay to reflect pulse storage duration. As depicted in Figure 3-(a) to (f), the inner waveguide width is varied from 0.35 to 0.45 µm across the series to demonstrate its effect, while the periodicity and width of the corrugated waveguide are adjusted to highlight their impact on the time delay and peak intensity. The process employs a figure of merit (FoM) based on a time-domain monitor. This FoM is derived by multiplying the normalized time delay by the normalized peak intensity. This methodology is favored over the traditional approach, which relies on the delay-bandwidth product, particularly for selecting structural parameters based solely on time-domain features. The rationale behind this preference is the directly proportional relationship between the broadband characteristics of the structure and the peak intensity of the output pulse \cite{agrawal2013nonlinear}. Consequently, identifying the maximum points from this multiplication allows us to determine the optimal structural parameters that achieve excellent performance in terms of peak power, temporal resolution, and pulse storage duration simultaneously.

In Figure 3-(a), it is observed that in the region marked by gold and black stars, an increased alteration in the delay amount leads to a reduction in peak intensity by up to 17\%. This result emphasizes the importance of designing the grating structure in the time-domain rather than focusing solely on frequency domain characteristics like achieving high group velocity dispersion. In the analysis of Figure 3-(b), it is evident that despite the constancy of the time delay level, there is a potential increase in the peak intensity level. This observation also emphasizes the critical importance of precisely selecting each parameter for optimal ultrafast propagation performance. In Figure 3-(c) to (e), it is noted that for inner waveguide width ($w_1$) values of 390 nm, 400 nm, and 420 nm, respectively, there is a decrease in time delay following each expansion, which correspondingly results in a gradual increase of the peak intensity. As illustrated by the white star in Figure 3-(f), elevating the $w_1$ parameter to 450 nm results in a peak E-field intensity of 0.64 V/m, yielding a delay of 0.16 ps in a structure with 50 periods. When comparing the optimal points in Figure 3-(a) and (f), there is a 33\% decrease in time delay and a 21\% increase in the peak intensity value. 

\begin{table}[!ht]
\renewcommand{\arraystretch}{1.3} 
\caption{List of the Parameters for selected 1DGWs
(1DGW\#1: High Delay-Oriented Structure, 
1DGW\#2: Time-Domain-Oriented Structure)}
\label{tab:table1}
\centering
\begin{tabular}{lccc}
\hline
& \textbf{1DGW \#1} & \textbf{1DGW \#2} \\
\hline
DC [\%] & 50 & 50 \\
\(w_1\) [nm] & 350 & 450 \\
\(w_2\) [nm] & 2150 & 2270 \\
\(w_3\) [nm] & 2050 & 2170 \\
\(\Lambda\) [nm] & 390 & 365 \\
\# of Periods & 75, 150 & 150 \\
\hline
\end{tabular}
\end{table}

Continuing to increase the $w_1$ value beyond 450 nm becomes less efficient because exceeding this value causes the corrugation width to decrease and shifts the operational mode of the structure to reflection, as shown in Figure 2. This shift substantially reduces the attainable delay at 1550 nm and requires a significant increase in the number of periods to achieve the optimal pulse storage duration. Therefore, 450 nm is selected as the upper limit for $w_1$. To highlight the significance of the proposed ultrafast time-domain-oriented design, we select two specific designs: the high delay-oriented design, marked by a black star in Figure 3-(a), and the ultrafast time-domain-oriented design, indicated by a white star in Figure 3-(f). The high delay-oriented design, when compared to the ultrafast time-domain-oriented design, shows a 63\% increase in time delay and a 31\% reduction in peak intensity. Figure 3-(g) and (h) illustrate how the time delay and peak intensity values vary with respect to the corrugated grating width ($w_2$) at the period values ($\Lambda$) of the specifically chosen designs. Especially considering the fabrication constraints, the parameters closest to the chosen values and compatible with the fabrication conditions are determined within the green acceptable regions, as shown in Figure 3-(g) and (h). 
The list of final parameters for the chosen structures, represented by the black and white stars in Figure 3-(a) and (f), is provided in Table 1. 1DGW\#1 (black, as shown in Figure 3-(a)) and 1DGW\#2 (white, as shown in Figure 3-(f)) represent design parameters for high-delay and ultrafast time-domain-oriented structures, respectively.

\begin{figure*}[!t]
\centering
\includegraphics[width=6.5in, height=4.9in]{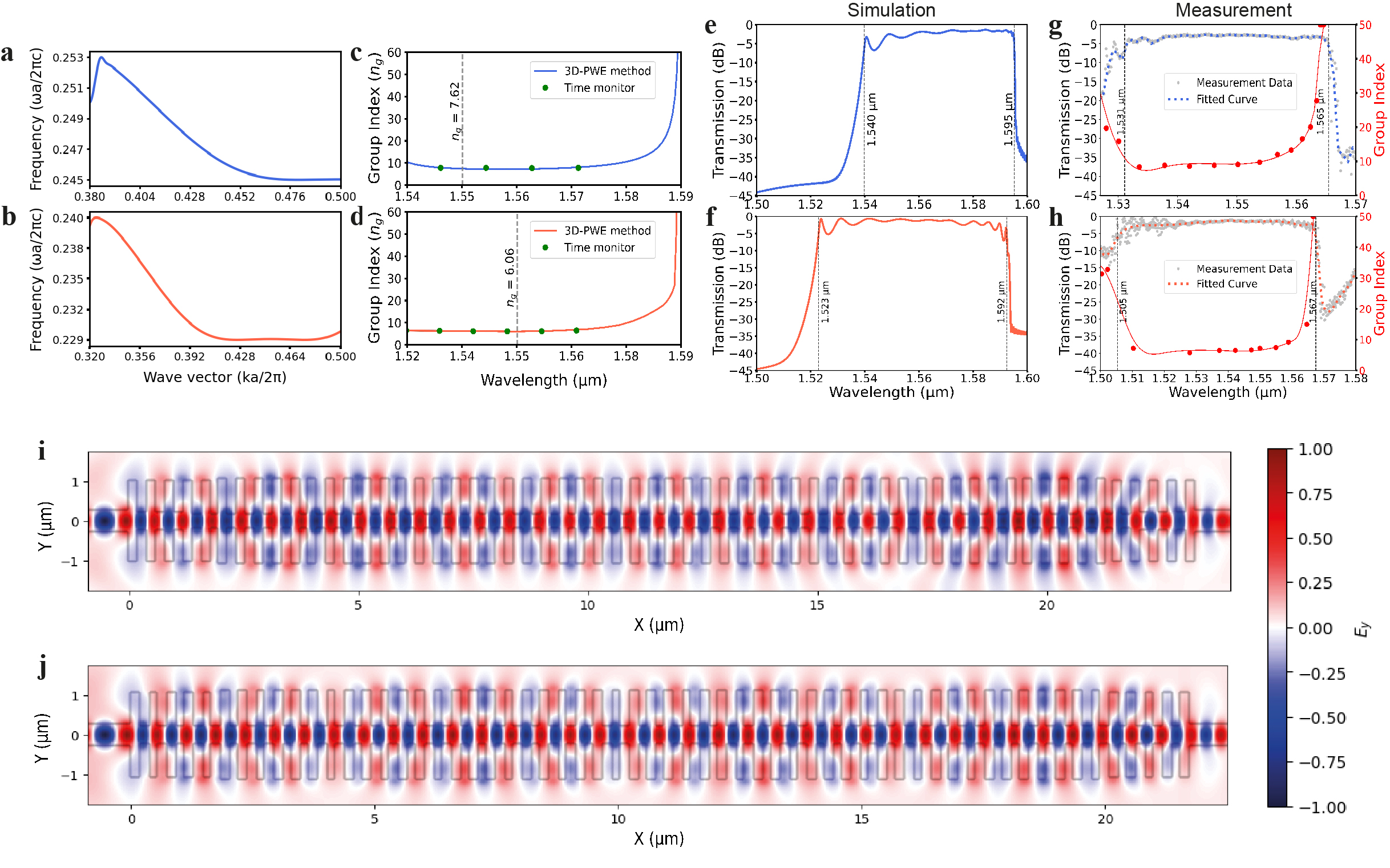}
\caption{(a)-(b) Dispersion relation of the fundamental quasi-TE Bloch mode for two specialized 1DGWs, the blue curve represents 1DGW\#1, while the orange curve corresponds to 1DGW\#2, (c)-(d) group index results obtained from 3D-PWE and 3D-FDTD simulations for 1DGW\#1 (blue) and 1DGW\#2 (orange) structures, respectively, (e)-(f) simulated transmission spectra for 1DGW\#1 (blue) and 1DGW\#2 (orange), with the 3dB bandwidth boundaries, (g)-(h) measured transmission and group index data for 1DGW\#1 (blue) and 1DGW\#2 (orange), respectively, (i)-(j) simulated steady-state $E_{y}$ intensity distributions at 1550 nm for 1DGW\#1 and 1DGW\#2, respectively.}
\label{s4}
\end{figure*}

\begin{figure*}[!ht]
\centering
\includegraphics[width=6.5in,height=6.5in]{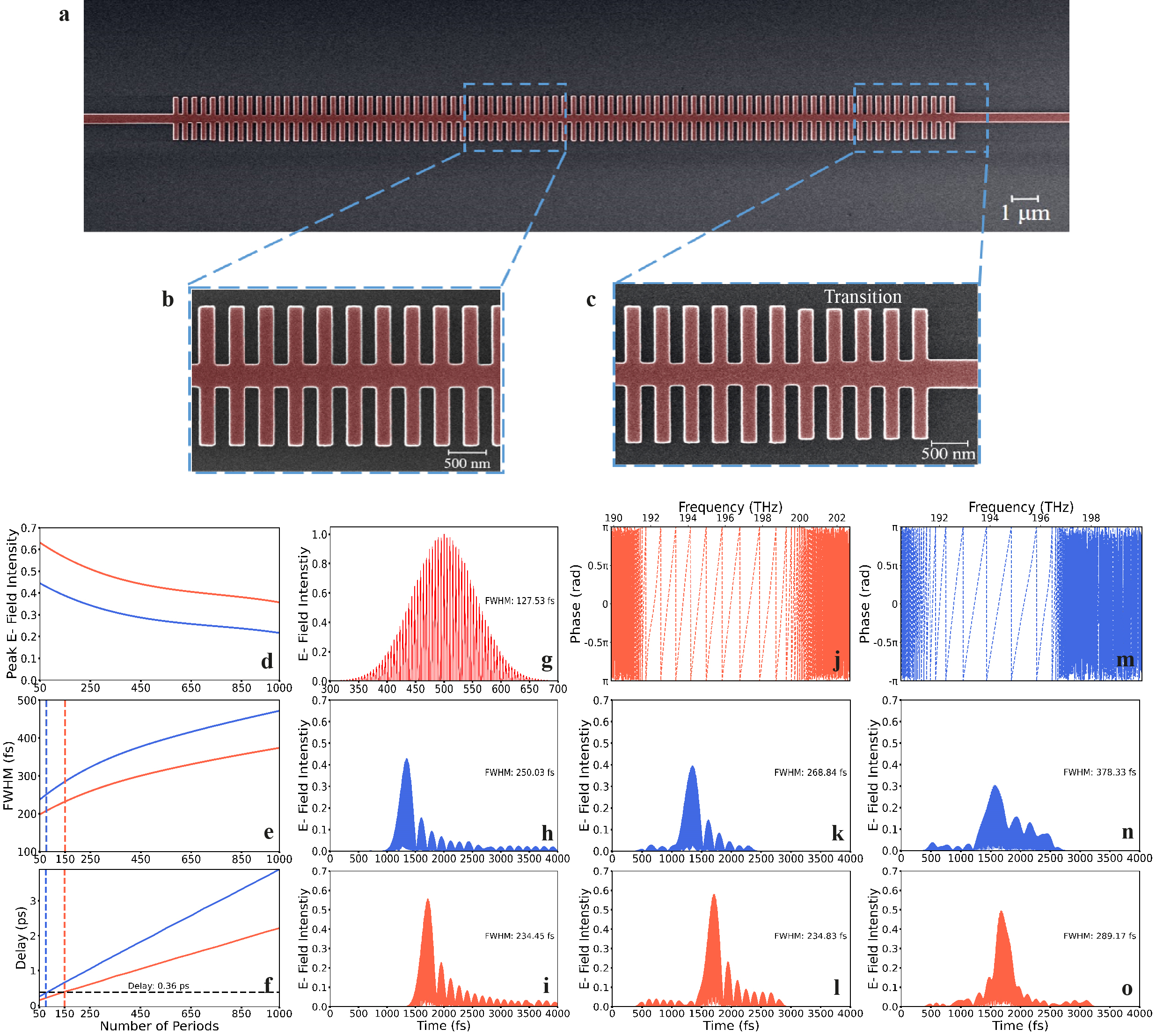}
\caption{(a) Colored SEM image of the 1DGW structure, (b) close-up SEM image of the 1DGW periodic region, (c) close-up SEM image of the transition region, (d) peak E- field intensity variation as a function of the number of periods, (e) FWHM variation as a function of the number of periods, (f) change in time delay as a function of the number of periods, (g) input E-field signal  (FWHM value of the power Gaussian signal is 90 fs), (h) simulated time-domain output for the 1DGW\#1 structure, (i) simulated time-domain output for the 1DGW\#2, (j) phase change as a function of frequency for the 1DGW\#2 structure, (k) time-domain output for 1DGW\#1 structure using FIR modeling with simulation-based S-parameters, (l) time-domain output for 1DGW\#2 structure using FIR modeling with simulation-based S-parameters, (m) phase change as a function of frequency for the 1DGW\#1 structure, (n) time-domain output for 1DGW\#1 structure using FIR modeling with measurement-based S-parameters, (o) time-domain output for 1DGW\#2 structure using FIR modeling with measurement-based S-parameters.}

\label{s5}
\end{figure*}

 Figure 4-(a) and (b) exhibit the bandstructures for two selected topologies associated with quasi-transverse electrical (quasi-TE) modes. Their dispersion behavior can be described by the equation $k(\omega) = n_B(\omega)\frac{\omega}{c}$. Here, $k(\omega)$ represents the wavevector, $n_B(\omega)$ denotes the Bloch effective index, $\omega$ is the angular frequency, and $c$ is the speed of light in the medium. The group index of structures can be determined from its bandstructure by employing the formula $n_g = \frac{c}{\partial k / \partial \omega}$, where $n_g$ represents the group index.  The three-dimensional plane-wave expansion (3D-PWE) method is utilized to obtain bandstructures, theoretically enabling the achievement of an infinite group index and local density of states (LDOS) due to the unlimited extent of the simulated geometry \cite{jean2019slow}. When employing 3D-FDTD simulations with finite geometries, as shown in Figure 4-(c) and (d), a notable limitation emerges: beyond a certain threshold, the increased decay rate of the ultrafast pulse makes it impractical to determine the group index from the time monitor \cite{tsakmakidis2021topological}.

\section{Device Fabrication and  Experimental Verification}
\subsection{Fabrication}

The designed devices are fabricated using the NanoSOI Multi-Project Wafer (MPW) process by Applied Nanotools Inc., employing direct-write 100 keV electron beam lithography (EBL) \cite{chrostowski2019silicon,canadian_foundry_2024}. This process utilizes 8-inch SOI wafers with a 220 nm thick device layer and a 2 $\mu$m thick buffer oxide layer. The photonic devices are patterned using a JEOL JBX-8100FS EBL system. The designed devices are etched to the buffer oxide layer using an anisotropic inductively coupled plasma-reactive ion etching (ICP-RIE) process. Figure 5-(a) to (c) showcase a colored scanning electron microscope (SEM) image of the device. After taking the SEM image, a 2.2 $\mu$m thick oxide cladding is deposited using a plasma-enhanced chemical vapor deposition (PECVD) process.

\subsection{Experimental Verification}

The measurements are conducted using an automated test setup provided by the SiEPIC program test service \cite{chrostowski2019silicon, canadian_foundry_2024}. Light emitted by a tunable laser (Agilent 81600B), with a wavelength sweeping incrementally from 1480 nm to 1580 nm in 8 pm steps, first passes through a polarization controller to ensure TE polarization. It is then coupled to the input grating coupler via a polarization-maintaining fiber (PMF). The guided light propagates through the 1DGWs and reaches the output grating couplers, where it is collected by a single-mode PMF and subsequently directed into optical power sensors (Agilent 81635A). The transmission measurements include losses associated with the input/output grating couplers, bends, and waveguides. Moreover, the bandwidth of the 1DGWs can only be accurately determined after addressing the bandwidth limitations imposed by the grating couplers. To normalize the transmission values, a baseline measurement is conducted on a strip waveguide equipped with grating couplers and an identical number of bends as the characterized 1DGW systems. The effects outside the device under test (DUT) are eliminated using the average transmission spectra obtained from ten different loopback structures.

\section{RESULTS AND DISCUSSION}

\subsection{Insertion Loss and 3 dB Bandwidth}

The transmission results obtained from simulations for the designed 1DGW structures are depicted in Figure 4-(e) and (f), while the corresponding experimental results are illustrated in Figure 4-(g) and (h). The simulation results presented in Figure 4-(e) for 1DGW\#1 demonstrate that the 3-dB bandwidth is approximately 55 nm and the observed insertion loss at a wavelength of 1.55 \(\mu\)m  is 2.6 dB. The 3-dB bandwidth, as determined from measurement, is 34 nm with an insertion loss of 2.65 dB at 1.55 \(\mu\)m, as depicted in Figure 4-(g). The narrowing of the bandwidth by approximately 21 nm is attributed to fabrication effects, primarily arising from dimensional variations such as sidewall roughness, period, and corrugation width during the lithography process \cite{saghaei2022sinusoidal}. As detailed in Section II, the Bragg wavelength ($\lambda_B$) is directly related to the grating period ($\Lambda$), meaning that even slight deviations in the duty cycle (DC) or period can result in significant shifts in the device's performance characteristics \cite{wang2012narrow}. For wavelength drift calculation, the formula \(\Delta\lambda = \lambda_{\text{center\_measured}} - \lambda_{\text{center\_simulated}}\) is employed. With the center frequencies of the 3 dB bandwidths in simulation and measured data calculated as $\lambda_{\text{center\_simulated}}$ equal to 1567 nm and $\lambda_{\text{center\_measured}}$ equal to 1548 nm, the resulting wavelength drift is 19 nm. This calculation yields a drift wavelength ratio of \(1.21\%\), highlighting the impact of fabrication variations. In the case of the 1DGW\#2 structure, the analysis provides a 3 dB bandwidth of 69 nm from simulations, alongside an insertion loss of 1.13 dB at 1.55 \(\mu\)m as shown in Figure 4-(f). The insertion loss recorded from measurement results is 1.49 dB at 1.55 \(\mu\)m, within a narrower 3 dB bandwidth of 62 nm, as displayed in Figure 4-(h). This reflects a decrease of 7 nm in the 3 dB bandwidth from the simulated to the measured results. Using the previously described formula, calculating the wavelength drift ratio results in an approximate drift of 21 nm, equivalent to a drift ratio of \(1.35\%\). Consequently, according to the measurement results, the time-oriented design exhibits a 43.77\% lower loss and an 82.35\% wider 3dB bandwidth compared to the high group index-based design.

The differentiation of losses can also be elucidated by examining the mode profiles. As illustrated in Figure 4-(i) and (j), the spatial distribution of the $E_{y}$ mode profiles is depicted for the 1DGW\#1 and 1DGW\#2 structures, respectively. When examining the mode profiles of 1DGW\#1 and 1DGW\#2, particularly in the transition regions, an increase in scattering is observed for the 1DGW\#1 structure, despite both structures having identical transition concepts. This difficulty arises from the challenge of achieving a smooth transition from a high group index dispersion profile (e.g., the 1DGW\#1) to a lower group index dispersion profile (e.g., strip waveguide) with a finite transition design, as illustrated in Figure 4-(c). Despite both structures utilizing the same anti-Fresnel reflection method-based, high-bandwidth transition design, the design of 1DGW\#2 demonstrates greater robustness against performance variations in the transition region, which can be significantly affected by fabrication imperfections. This enhanced performance is attributed to the proposed time-domain-oriented design methodology, which incorporates the effects of the transition region.

\subsection{Group Index and Slow Light Bandwidth}

Following the transmission results, the group index and slow light bandwidth are demonstrated using simulation data in Figure 4-(c) and (d), along with the measurement results in Figure 4-(g) and (h). Determining slow light bandwidth (\( \Delta \lambda_{SL} \)) requires identifying the boundaries of the slow light regions. These boundaries are defined as the wavelengths where the group index is within 10 percent of the group index measured at a wavelength of 1.55 \(\mu m\). Conducting a post-fabrication analysis of the group index is also essential, as group index values can be significantly affected by imperfections in the transition zone and various other fabrication-related factors. The measured group index for the 1DGWs is determined by analyzing the interference pattern observed in the spectrum of an unbalanced Mach-Zehnder interferometer (MZI) test structure. The first MZI consists of one arm with a 62 \(\mu\)m long grating waveguide and the other arm featuring a 62 \(\mu\)m long, 500 nm wide silicon single-mode strip waveguide, used for calculating the group index of the 1DGW\#1 design. Similarly, for the 1DGW\#2 design, these lengths change slightly to 58 \(\mu\)m for both arms due to the unit length difference. The group indices of the 1DGW structures are calculated based on the measured transmission results of the MZI structures, as described in Equation (3) \cite{torrijos2021slow}.

\begin{equation}
n_{g,1DGW} = n_{g,strip} + \frac{\lambda_{min} \times \lambda_{max}}{2 \times L \times (\lambda_{min} - \lambda_{max})}
\end{equation}

\noindent Here, \( n_{g, 1DGW} \) represents the group index of the arm containing the 1DGW, while \( n_{g, \text{strip}} \) refers to the group index of the TE mode strip waveguide, which has been experimentally determined to be approximately 4.11. The parameter \( L \) denotes the physical length of the 1DGW, and \( \lambda_{\text{max}} \) and \( \lambda_{\text{min}} \) indicate the wavelengths corresponding to the local power maximum and minimum, respectively.

As illustrated in Figure 4-(c), for 1DGW\#1, the simulated group index \( n_g \) at a wavelength of 1.55 \(\mu m\) is determined to be 7.62, with a corresponding slow light bandwidth (\( \Delta \lambda_{SL} \)) of 29 nm, spanning the wavelength range of 1546 to 1575 nm. Similarly, as illustrated in Figure 4-(d), for 1DGW\#2, which is specifically designed for the propagation of ultrafast pulses, the simulated group index is calculated as 6.06, with a \( \Delta \lambda_{SL} \) of 34 nm, covering a wavelength range from 1531 to 1565 nm. Considering the experimental results, Figure 4-(g) and (h) reveal that the group indices at a wavelength of 1.55 \(\mu\)m are 9.06 and 7.19 for 1DGW\#1 and 1DGW\#2, respectively. For the measured slow light bandwidth values, it is observed that the slow light bandwidth \( \Delta \lambda_{SL} \) for the structure 1DGW\#1 spans 22 nm, ranging from 1532 nm to 1554 nm. For the structure 1DGW\#2, the slow light bandwidth \( \Delta \lambda_{SL} \) of 21 nm is recorded, extending from 1536 nm to 1557 nm. When comparing the measured and simulated results, a decrease of 7 nm in the slow light bandwidth \( \Delta \lambda_{\text{SL}} \) for 1DGW\#1 and 13 nm for 1DGW\#2 is observed. Changes in the slow light bandwidth and group index are primarily attributed to shifts in the center wavelength and imperfections caused by fabrication processes. These factors significantly contribute to the observed variations in the slow light bandwidth between simulation and experimental data \cite{hao2019increasing}. Despite the wavelength drift and deviations in the slow light region, the overall trend aligns well with the simulated expectations.

\subsection{Ultrafast Pulse Propagation Time-Domain Dynamics in 1DGWs}

After determining the transmission and group index values, which are essential for obtaining the scattering parameters for the FIR model, we present the time-domain analysis of the designed structure, based on both simulations and experiments, as shown in Figure 5-(d) to (o). In all graphs, the blue color represents the results of the 1DGW\#1 design, whereas the orange color corresponds to the results of the 1DGW\#2 design. The simulation results shown in Figures 5-(d) to (i) are obtained using the 3D-FDTD time monitor. In these simulations, a Gaussian signal is used as input, defined by a central wavelength of 1550 nm, a FWHM of 90 fs, and a time offset of 500 fs. As observed in Figure 5-(d), an increase in the number of periods leads to a great reduction in peak intensity. The proposed 1DGW\#2 structure demonstrates a peak intensity value of approximately 0.4 V/m over 1000 periods, as shown by the orange line. The peak intensity value decreases to approximately 0.2 V/m for the same number of periods in the 1DGW\#1 structure, as represented by the blue line. The peak intensity decay rate highlights the advantage of designing the structure with a focus on ultrafast time-domain characteristics.

Upon examining Figure 5-(e), it becomes apparent that the FWHM value expands as the number of periods increases. The rate at which the FWHM increases for the 1DGW\#2 structure is significantly less than that for the 1DGW\#1 structure. The lower pulse broadening value observed in the 1DGW\#2 structure can be attributed to the correlation between the peak intensity and the pulse broadening value. The trend similarity between Figure 5-(d) and (e) provides clear evidence supporting this statement. When comparing the FWHM values for 1000 periods, it is shown that the pulse broadening rate in the 1DGW\#2 structure is 26\% lower. However, to ensure a more accurate comparison, the designed structures should be evaluated based on an equal time delay criterion. Figure 5-(f) depicts the correlation between the number of periods and time delay. This visualization highlights the points at which the two specialized designs provide equivalent time delay values. Therefore, the target time delay is established as four times the FWHM of the input signal. It is observed that the 1DGW\#1 and 1DGW\#2 structures reach this delay at 75 and 150 periods, respectively, resulting in a time delay of 0.36 ps, as indicated by the dashed horizontal line in Figure 5-(f).

The applied input E-field signal, where the FWHM is observed to be 127 fs, can be seen in Figure 5-(g) (for power, this corresponds to 90 fs). Upon examining the time-domain outputs presented in Figure 5-(h) and (i) for selected specific time delay, it is observed that the blue pulse, representing the output from 1DGW\#1, has a peak intensity value of approximately 0.43 V/m and an FWHM value of 250 fs. Moreover, 1DGW\#2, engineered for enhanced performance in the ultrafast time-domain, is depicted in orange, as shown in Figure 5-(i). The pulse for 1DGW\#2 shows a peak E-field intensity value of 0.56 V/m and an FWHM of 234 fs.  Notably, the decrease in peak intensity value is significantly minimized during the same time delay value for 1DGW\#2. Moreover, the 1DGW\#2 structure achieved a 6\% reduction in pulse broadening compared to 1DGW\#1, thereby also reducing pulse distortion under the equal time delay condition. As previously noted, the primary focus of this design method is the peak power level and its associated temporal resolution. Therefore, the power-based comparison should be derived using the relationship \( I \propto E^2 \), where \( I \) is the peak power intensity and \( E \) is the E-field intensity. The FWHM of the power intensity follows \( \Delta t_I = \frac{\Delta t_E}{\sqrt{2}} \), where \( \Delta t_I \) and \( \Delta t_E \) represent the FWHM of the power and E-field, respectively. We should also note that, in this design method, we use the FWHM of the signals to evaluate and compare their temporal resolution. Consequently, in a power-based comparison, 1DGW\#2 demonstrates a 72\% increase in peak power and a 6\% smaller reduction in temporal resolution, as indicated by the decreased FWHM broadening, relative to 1DGW\#1.

To obtain the time-domain response of 1DGWs experimentally, the FIR model trained with complex and band-limited scattering parameters obtained from continuous wave-based measurement is employed. This approach is an effective and accurate method for analyzing ultrafast pulse propagation in passive, lossy, linear, and time-invariant (LTI) systems. This is because these structures, such as designed 1DGWs, are characterized by scattering parameters that adeptly account for nonidealities such as higher-order dispersion, wavelength-dependent losses, and inaccuracies in coupling coefficients \cite{ye2019time}. Hence, it can be seen that all the necessary information to obtain time-domain response can be acquired from classical optical measurements.

The FIR model is designed as the bandpass system, enabling faster results due to the absence of feedback-related pole residues, unlike in applications based on vector fitting \cite{gustavsen2009fast, ye2018numerical}. This modeling employs Lumerical Interconnect, a commercial software \cite{pond2014complete}. The software's capability to estimate modeling taps based on group delay, a feature intrinsic to the FIR modeling approach, enabled the realization of modelings with superior performance \cite{ansys_lumerical_2024}. When attempting to design with the Lumerical Interconnect IIR model based on the vector fitting algorithm using the same scattering parameters, numerous challenges are encountered, including difficulties in adjusting the parameters concerning the group delay and in determining the number of taps. Consequently, this approach yielded results that are more time-consuming and of lower resolution than those obtained with the FIR model. 

To evaluate the high-performance capabilities of the FIR approach, we initially designed an FIR model incorporating the scattering parameters of 1DGW\#1 and 1DGW\#2, utilizing Lumerical's 3D-FDTD simulation. The outcomes for 1DGW\#1 and 1DGW\#2 are depicted in Figure 5-(k) and (l) respectively. When Figure 5-(k) is analyzed, the peak E-field intensity is recorded at 0.39 V/m with an 8.62\% discrepancy from the 3D-FDTD results, and the FWHM is established at 268 fs, showing a 6.72\% difference compared to the 3D-FDTD findings. Moreover, upon examining Figure 5-(l), it is observed that the peak intensity achieves a value of 0.58 V/m, indicating a minimal discrepancy of 3.86\% compared to the 3D-FDTD result. Additionally, the FWHM value is determined to be 234 fs, which closely aligns with the result from the 3D-FDTD analysis. Therefore, it has been demonstrated that the error rates are significantly low. 

Following the same approach as explained above, we extract the measured time-domain response of the fabricated structures by employing the FIR modeling to the measured scattering parameters. To determine the phase data, the group delay is first calculated using the formula $\tau_g = \frac{L}{n_g / c}$ where  $\tau_g$ is the group delay. Subsequently, by employing Equation (4), the phase is calculated from the group delay, which itself is determined through linear interpolation of the group index points, utilizing the cumulative trapezoidal numerical integration technique \cite{yeh1978comparison}.

\begin{equation}
\tau_g = \frac{\Delta \phi}{\Delta \omega}
\end{equation}

\noindent Here, \(\Delta\phi\) denotes the change in phase throughout the observed frequency span \(\Delta w\). Upon examining Figure 5-(j) and (m), the phase versus frequency results for 1DGW\#2 and 1DGW\#1 can be observed, respectively, within the range of \(-\pi\) to \(\pi\). Utilizing the acquired phase data and transmission measurements, we derive the scattering parameters necessary for training the FIR model. Figure 5-(n) and (o) display the results of ultrafast signals transmitted through 1DGW\#1 and 1DGW\#2, respectively, using the trained FIR model. When analyzing the output for the 1DGW\#1 structure in Figure 5-(n), it is observed that the FWHM value is 378 fs, and the peak intensity value is approximately 0.3 V/m.  When comparing the simulation results with the experimental results, a 30\% decrease in peak intensity and a 51\% increase in FWHM expansion are observed for the 1DGW\#1 structure. Upon analyzing Figure 5-(o) for the 1DGW\#2 structure, it is observed that the peak intensity value has decreased to 0.50 V/m. The FWHM value of 1DGW\#2 is observed to be 289 fs.  Hence, for the 1DGW\#2 structure, the peak intensity value decreased by 11\%, and the FWHM expansion increased by 24\% compared to the simulation result. 

When comparing the experimental results of the two designs, the peak E-field intensity value of the time domain-oriented 1DGW\#2 structure is observed to be 66\% higher than that of 1DGW\#1. By focusing on power level metrics, based on the previously explained relationship between power and the E-field, the experimental measurements show that 1DGW\#2's peak power intensity is 2.78 times higher than that of 1DGW\#1. Furthermore, the broadening in the FWHM value is reduced by 24\% for 1DGW\#2 compared to 1DGW\#1. In a power signal-based comparison of FWHM, this reduction in broadening is also 24\%. This is a consequence of the higher dispersion profile-based design of 1DGW\#1, which makes it more sensitive to fabrication imperfections. Consequently, the 6\% difference observed in simulations for the FWHM has increased to 24\% in the measurements. In contrast, despite a slight narrowing of the 3 dB bandwidth and minor wavelength drift, the superior performance of the 1DGW\#2 structure can be attributed to its dispersion profile, optimized for time-domain operations, which results in more robust performance against fabrication imperfections. Therefore, it exhibits a 24\% smaller sacrifice in temporal resolution, as reflected by the decreased FWHM broadening, in comparison to 1DGW\#1.

\section{Conclusion}

This study presents a time-domain dynamics-based design method for dispersive integrated waveguides, with a specific focus on 1DGWs for ultrafast pulse propagation. This study primarily focuses on achieving high peak power and high temporal resolution in dispersive integrated waveguides for femtosecond technology applications. However, the balance between these factors can be adjusted for specific applications using the proposed design methodology. The implications of our proposed design approach are far-reaching, with immediate applications in temporal-mode encoding using highly dispersive structures for high-dimensional quantum information processing. Particularly, in systems utilizing ultrafast pulses to generate diverse quantum states, the reduction of pulse shape distortion becomes paramount for maintaining system fidelity. Furthermore, our study eases the challenge of measuring complex temporal waveforms using the FIR model approach, a long-standing obstacle in high-dimensional quantum information processing. Our approach also holds significant potential for various other applications, including photonic neuromorphic computing, photonic neural networks, photonic LIDAR, and terahertz waveform generation.

\section*{Author contributions}
All authors have accepted responsibility for the entire content of this manuscript and approved its submission.

\section*{Acknowledgement}
We acknowledge the edX UBCx Silicon Photonics Design, Fabrication, and Data Analysis course supported by the Natural Sciences and Engineering Research Council of Canada (NSERC) and Silicon Electronic-Photonic Integrated Circuits (SiEPIC) program.

\section*{Funding}
This research was supported by the Scientific and Technological Research Council of Turkey (TUBITAK), project number 122E566. Ahmet O. Sakin acknowledges the support of the TUBITAK BIDEB 2210A grant.

\section*{Conflict of interest}
The authors declare no conflict of interest.

\bibliographystyle{unsrt}  
\bibliography{references}

\end{document}